\documentclass[aps,twocolumn,longbibliography]{revtex4-1}
\usepackage[dvips]{graphics,graphicx}
\usepackage{amsmath}
\usepackage[usenames, dvipsnames]{color}
\usepackage{graphicx}
\usepackage{graphics}
\usepackage[outdir=./]{epstopdf}

\begin{document}

\title{Ballistic transport of interacting Bose particles in the tight-binding chain}
\author{ P. S. Muraev$^{1,2}$, D. N. Maksimov$^{1,3}$, and A. R. Kolovsky$^{1,2}$}
\affiliation{$^1$Kirensky Institute of Physics, Federal Research Center KSC SB RAS, 660036, Krasnoyarsk, Russia}
\affiliation{$^2$School of Engineering Physics and Radio Electronics, Siberian Federal University, 660041, Krasnoyarsk, Russia}
\affiliation{$^3$IRC SQC, Siberian Federal University, 660041, Krasnoyarsk, Russia}
\date{\today}
\begin{abstract}
It is known that quantum transport of non-interacting Bose particles across the tight-binding chain is  ballistic in the sense that the current does not depend on the chain length. We address the question whether the transport of {\em strongly interacting} bosons can be ballistic as well. We find such a regime and show that, classically, it corresponds to the synchronized motion of local non-linear oscillators. It is also argued that, unlike the case of non-interacting bosons, the transporting state responsible for the ballistic transport of interacting bosons is metastable, i.e., the current decays in course of time. An estimate for the decay time is obtained.
\end{abstract}
\maketitle

\section{Introduction}

In the past decade much efforts were invested in understanding quantum transport of Bose particles across the one-dimension lattice connecting  two particle reservoirs \cite{iv2013bosonic,kordas2015dissipative,kolovsky2018landauer,bychek2020open,muraev2022resonant}.
Several theoretical  approaches have  been used to analyze this problem, including the straightforward numerical
simulations of the master equation for bosons in the lattice, quantum jumps methods, and the semiclassical (mean field) and pseudoclassical approaches. The last two approaches are especially important for developing an intuitive physical picture because they map the quantum transport problem to the classical problem of excitation transfer in a chain of coupled nonlinear oscillators with the edge oscillators driven by external forces, where the type of the driving force is determined by the ergodic properties of the particle reservoirs.  Namely, if reservoirs justify the Born-Markov approximation, the edge oscillators are driven by the complex white noise whose intensity is proportional to the particle density in reservoir
  \cite{iv2013bosonic,kordas2015dissipative,kolovsky2018landauer}.  For non-Markovian reservoirs the white noise has to be superseded   by the narrow-band noise with the spectral density spanning a finite frequency interval \cite{bychek2020open}.  Typically, this is the case where Bose particles in reservoirs are close to condensation.  At last, one may consider the   situation where the spectral density of the colored noise is given by the  $\delta$-function, i.e., we have a periodic driving.  Experimentally, this case is realized, for example, in the chain of the capacitively coupled transmons where the first  transmon is excited by a microwave generator \cite{raftery2014observation,fitzpatrick2017observation,fedorov2021photon},  or in the array of optical cavities with the Kerr nonlinearity where the first cavity is excited by a laser.  We mentioned that the minimal size chains consisting of two cavities  is currently used to study a number of other   fundamental problems
     \cite{lagoudakis2010coherent,abbarchi2013macroscopic,cao2016two,casteels2017quantum,rodriguez2017probing,lledo2020dissipative,zambon2020parametric}.  In the present work, however, we focus exclusively  on the transport problem where the main question is the current of Bose particles across the chain. As the main result, we show that the   edge-driven systems can exhibit the exotic transport regime where the current of {\em strongly interacting}  bosons is independent of the chain length and is insensitive to a weak disorder. This relates the reported   results to the problem of super-fluidity of Bose gases \cite{astrakharchik2004motion,cherny2012theory}.

\section{The model}
\label{sec2}

In the rotating wave approximation the quantum  Hamiltonian of the system under scrutiny has the form
\begin{eqnarray}
\nonumber
\widehat{{\cal H}}= \sum_{\ell=1}^{L}\hbar(\omega_\ell-\nu)\hat{n}_{\ell}
-\frac{\hbar J}{2}\left(\sum_{\ell=1}^{L-1}\hat{a}_{\ell+1}^{\dagger}\hat{a}_{\ell} +{\rm h.c.} \right)  \\
\label{a1}
+\frac{\hbar^2U}{2}\sum_{\ell=1}^{L}\hat{n}_{\ell} (\hat{n}_{\ell}-1) + \frac{\sqrt{\hbar}\Omega}{2}(\hat{a}_1^\dagger + \hat{a}_1) \;,
\end{eqnarray}
where the index $\ell$ labels the chain site, $\hat{a}_\ell$ and $\hat{a}^\dagger_\ell$ are the creation and annihilation  bosonic operators commuting to unity,  $\hat{n}_\ell=\hat{a}^\dagger_\ell\hat{a}_\ell$ is the number operators,  $\omega_\ell$  are the linear frequencies (on-site energies),  $J$ is the hopping matrix element, $U$ the interaction constant (nonlinearity), and the Rabi frequency $\Omega$ characterizes the strength of the external monochromatic driving with the frequency $\nu$. We shall denote the detuning $\nu-\omega_\ell$ by $\Delta_\ell$ where the absence of the subindex $\ell$ will imply identical  on-site energies.

Assuming only the last oscillator is subject to decay, the governing master equation reads
\begin{equation}
\label{a2}
\frac{\partial \widehat{{\cal R}}}{\partial t}=-\frac{i}{\hbar}[\widehat{{\cal H}},  \widehat{{\cal R}}] -\frac{\gamma}{2}
\left(\hat{a}_L^{\dagger}\hat{a}_L\widehat{\cal R }-2\hat{a}_L\widehat{\cal R }\hat{a}_L^{\dagger}
+\widehat{\cal R }\hat{a}_L^{\dagger}\hat{a}_L \right) \;,
\end{equation}
where $\gamma$ is the relaxation constant. We mention in passing that the results reported below also hold true in the case where the other  oscillators are also subject to decay but their decay rates $\gamma_\ell \ll \gamma$. To address  the quantum-to-classical correspondence, we incorporate in the Hamiltonian (\ref{a1}) and the master equation (\ref{a2}) the effective Planck constant  $\hbar$, the physical meaning of which will be explained in the beginning  of Sec.~\ref{sec5}.

Our main object of interest is the single-particle density matrix (SPDM)
\begin{equation}
\label{a3}
\hat{\rho}(t)={\rm Tr}[\hat{a}^\dagger_\ell \hat{a}_m {\cal R}(t)] \;.
\end{equation}
The diagonal elements of this matrix give the occupation numbers of the chain sites while
the sub-diagonal determines the current across the chain
\begin{equation}
\label{a5}
 j(t)=\frac{1}{L-1}{\rm Tr}[\hat{j}\hat{\rho}(t)]  \;,
 \end{equation}
where $\hat{j}$ is the single-particle current operator with the elements $j_{\ell,\ell'} =J(\delta_{\ell,\ell'+1} -\mathrm{h.c.})/2i$. At the same time, as it follows from the continuity equation, the stationary current $\bar{j}=j(t\rightarrow\infty)$ is given by the population of the last site multiplied by $\gamma$, i.e., $\bar{j}=\gamma |a_L |^2$.

\section{Semiclassical analysis}
\label{sec3}

The semiclassical approximation  associates the mean values  of the creation and annihilation operators times $\sqrt{\hbar}$ with  the complex conjugated canonical variables $a_\ell$ and $a_\ell^*$. Then the governing equations take the form
\begin{eqnarray}
\nonumber
i\dot{a}_1=(-\Delta+U |a_1|^2) a_1-\frac{J}{2}a_2 +\frac{\Omega}{2} \;, \\
\nonumber
i\dot{a}_\ell=(-\Delta+U |a_\ell|^2) a_\ell-\frac{J}{2}(a_{\ell+1} +a_{\ell-1}) \;,\\
\label{b0}
\dot{a}_L=(-\Delta+U |a_L|^2) a_L-\frac{J}{2}a_{L-1} -i\frac{\gamma}{2} a_L \;.
\end{eqnarray}
Due to contraction of the phase volume  for $\gamma\ne0$, an arbitrary trajectories ${\bf a}(t)$ evolves to some attractor in the multidimensional phase space of the system. In what follows we focus on attractors which ensure the ballistic transport of  excitations from the first to the last oscillator. We begin with the case of vanishing inter-particle interaction where the system has a single attractor  -- a simple focus.  

\subsection{Vanishing inter-particle interaction}

\begin{figure}
\includegraphics[width=8.0cm,clip]{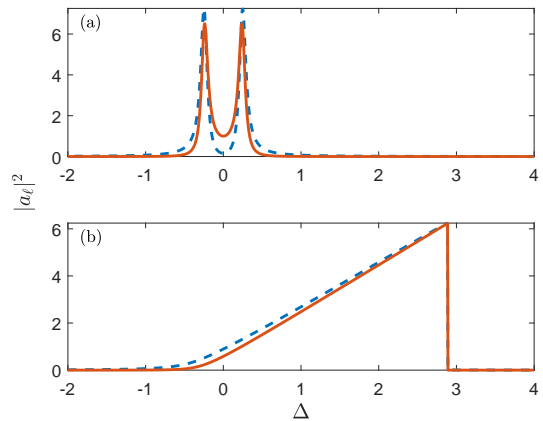}
\caption{Stationary values of the squared  amplitudes in the chain of the length $L=2$ as a function of the detuning $\Delta$ for $U=0$ (top), and $U=0.5$ (bottom). The other parameters are $J=0.5$, $\gamma=0.2$, and $\Omega=0.5$. The current across the chain is proportional to $|a_2|^2$ depicted by the red solid line.}
\label{fig1}
\end{figure}

For $U=0$ the system of coupled differential equations (\ref{b0}) can be decoupled by introducing the new canonical variables given by the eigenmodes  $X_\ell^{(j)}$ of the undriven ($\Omega=0$) chain. Since we excite the first oscillator  and the stationary current is proportional to the squared amplitude of the last oscillator, we have
\begin{equation}
\label{b1}
\bar{j} \sim  \left | \sum_{n=1}^L \frac{X_1^{(n)} X_L^{(n)} }{\Delta - \epsilon_n}  \right |^2 \;,
\end{equation}
where $\epsilon_n$ are the chain complex eigenfrequencies with
${\rm Re}[\epsilon_n]\approx -J\cos(\pi n/L)$ and ${\rm Im}[\epsilon_n]\sim \gamma$. It follows from Eq.~(\ref{b1}) that the stationary current as a function of the detuning shows $L$ peaks in the interval $|\Delta| <J/2$ --- the phenomenon known as the resonant transmission. The resonant transmission is illustrated in Fig.~\ref{fig1}~(a) for $L=2$. If $|U|\ll J$, the transmission peaks slightly bend to the left or right, depending on the sign of $U$. However, with a further increase of the interaction constant, the discussed simple attractor show a cascade of bifurcations \cite{giraldo2022semiclassical}, leading to a number of qualitatively different transport regimes.  We also would like to mention that the resonant transmission Eq.~(\ref{b1})   is sensitive to the on-site  disorder $\omega_\ell$ due to the presence of the product   $X_1^{(n)} X_L^{(n)}$ in Eq.~(\ref{b1}), which tends to zero in the regime of Anderson's localization.

\subsection{Strong inter-particle interaction}

\begin{figure}
\includegraphics[width=5.5cm,clip]{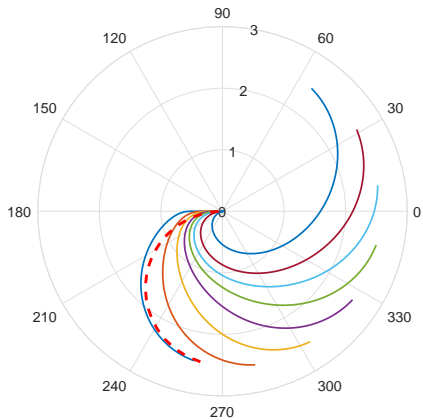}
\caption{The stationary complex amplitude of the local oscillators in the chain of the length $L=8$ for $-4<\Delta<2.5$. The dashed line is Eq.~(\ref{b3}). The other system parameters are as in Fig.~\ref{fig1}.}
\label{fig2}
\end{figure}

Next, we address the case $| U | > J$  and, to be specific, we shall consider positive $U$ from now on. In this case the attractor, which ensures  the ballistic transport, corresponds to the synchronized motion of the oscillators,
\begin{equation}
\label{b2}
a_{\ell+1}\approx a_{\ell} e^{i\phi} \;,\quad \phi\approx\arcsin(\gamma/J) \;.
\end{equation}
Equation (\ref{b2}) is illustrated in Fig.~\ref{fig2}  for $L=8$. The crucial feature of the transporting state (\ref{b2}) is the existence of the critical detuning $\Delta_{cr}$ above which the basin of the discussed attractor shrinks to zero.

Let us discuss the results shown in Fig.~\ref{fig1}(b) in more detail.  First, we notice that in the interval $0<\Delta<\Delta_{cr}$
 the squared amplitudes $|a_\ell |^2$ grow approximately linear with the detuning, i.e.,  $|a_\ell |^2 \approx \Delta/U$. For $\gamma=0$ this linear dependence exhibits the phenomenon of capturing in the nonlinear resonance \cite{lichtenberg2013regular}. In the presence of dissipation, however, the nonlinear resonance degenerates into the limit cycle. This transformation of the nonlinear resonance into the limit cycle can be studied in full details for $L=1$, i.e, for the dissipative driven nonlinear oscillator. In that system the stationary amplitude of the oscillator is given by the relation  \cite{landau1976mechanics},
\begin{equation}
\label{b3}
a=\frac{\Omega\slash 2}{U|a|^2 -\Delta-i\gamma/2} \;,
\end{equation}
where  $|a|^2$  obeys the algebraic equation
\begin{equation}
\label{b4}
|a|^2=\frac{(\Omega\slash 2)^2}{( U |a|^2 - \Delta)^2+(\gamma/2)^2} \;.
\end{equation}
We found that Eq.~(\ref{b3}) and Eq.~(\ref{b4}) provide a good approximation for the amplitude of the first oscillator in the chain if $L>1$, see dashed line in Fig.~\ref{fig2}.  Thus, we can use Eq.~(\ref{b4}) to obtain an estimate for $\Delta_{cr}$,
\begin{equation}
\label{b5}
\Delta_{cr} \approx  U(\Omega/\gamma)^2 \;.
\end{equation}
It is seen in Fig.~\ref{fig1}(b) that, when we exceed this critical value, the amplitude of the last oscillator in the chain drops  almost to zero, which results in the abrupt decrease of the  current.

\subsection{Basin size}

\begin{figure}
\includegraphics[width=8.0cm,clip]{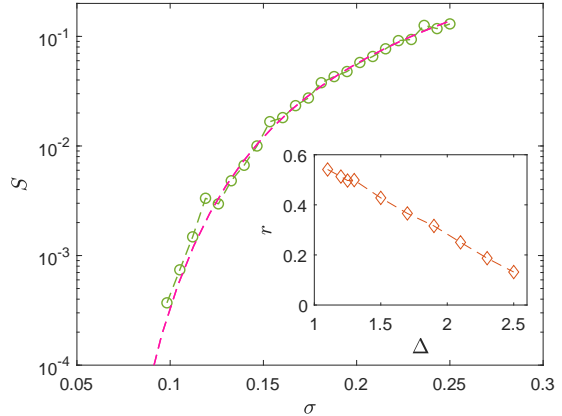}
\caption{ Main panel: Approximation of the numerical data by the function (\ref{b6}) for $\Delta=1.90$. Inset: The basin size as a function of the detuning $\Delta$.  The system parameters are the same as in Fig.~\ref{fig1}.}
\label{fig3}
\end{figure}

For future purposes we need to know the basin of the discussed attractor. Although the visualizing the attractor basin in the multi-dimensional phase space is difficult, one  can easily estimate its size \cite{shrimali2008nature}. To do this we randomly perturbed the stationary amplitude of the last oscillator as $a_L\rightarrow a_L+\xi$, where $\xi$ samples the Gaussian distribution with the width $\sigma$, and checked whether the perturbed trajectory attracts back to the solution (\ref{b2}).  Approximating the attractor basin by the circle (more precisely,  the basin projection on the $a_L$-plane)  we expect that the number of  not-attracted trajectories grows with the increase of $\sigma$ as
\begin{equation}
\label{b6}
S \sim \sigma^2 \exp\left(\frac{-r^2}{2\sigma^2}\right) \;,
\end{equation}
where $r$ is the circle radius.  Next, interpolating the numerical data by the function (\ref{b6}) we find $r=r(\Delta)$, see Fig.~\ref{fig3}. It is seen in Fig.~\ref{fig3} that the basin size decreases approximately  linearly with $\Delta$.

\subsection{Adiabatic passage}

We conclude this section by a remark that the results presented in Fig.~\ref{fig1} and Fig.~\ref{fig2}  can be fairly reproduced by using the adiabatic passage where the  detuning $\Delta$ is slowly changed in time. For the figures parameters we found no difference between the stationary and quasi-stationary solutions if the sweeping $\beta$,
\begin{displaymath}
\beta={\rm d}\Delta/{\rm d}t \;,
\end{displaymath}
is smaller than 100 tunneling periods $T=2\pi/J$ per unit interval of $\Delta$. It should be also stressed that, since we chose $U>0$,  we consider positive $\beta$. If the sweeping direction was  inverted, we would observe very different dynamical regimes, including the limit cycle in the frequency interval $0.37<\Delta<0.78$ ($L=2$) where the oscillator amplitudes periodically change in time.

\section{Quantum dynamics}
\label{sec4}

In this section we compare the results of the semiclassical analysis with the solution of the master equation (\ref{a2}).
We solve the master equation in the Hilbert space given by the direct sum of the subspaces associated with the fixed number
of particles in the chain, $N=0,1,\ldots,N_{max}$ where $N_{max}$ is the truncation parameter. We control the accuracy by
checking the convergence of the results as $N_{max}$ is increased.

\begin{figure}
\includegraphics[width=8.0cm,clip]{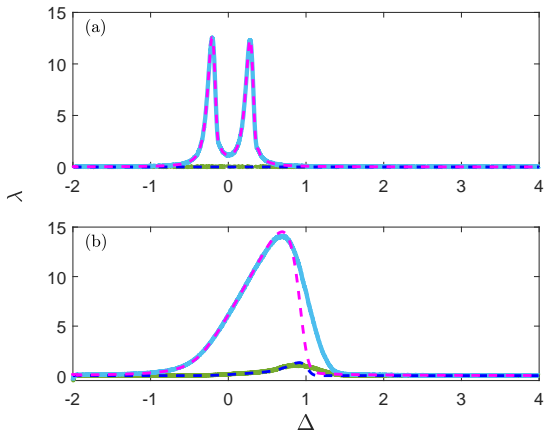}
\caption{Eigenvalues of the SPDM for $U=0$ and $\hbar=1$ (top) and  $U=0.5$ and $\hbar=0.25$ (bottom). The sweeping rate of the detuning $\Delta$ is $2\pi\cdot800/6$ per unit interval of $\Delta$. The dashed and solid lines are the exact result and the result of the pseudoclassical approach (average over 3600 realizations), respectively. }
\label{fig4}
\end{figure}

First, we study the transporting state of the system for $U=0$. We find  this state by sweeping the detuning $\Delta$  with a fixed rate $\beta$ in the interval $|\Delta|\gg J$. We take precaution that the rate $\beta$ is small enough to insure the adiabatic regime. The upper panel in Fig.~\ref{fig4} shows eigenvalues $\lambda_n=\lambda_n(\Delta)$ of the stationary SPDM of the system with $L=2$ sites.  Notice that the matrix has only one non-zero eigenvalue and this holds true for arbitrary $L$. Comparing the result shown in Fig.~\ref{fig4}(a) with  the result of the semiclassical analysis  we conclude that the stationary SPDM  is determined by the stationary solution ${\bf a}(t\rightarrow\infty)$ of the classical Eqs.~(\ref{b0}) through the relation $\bar{\rho}_{\ell,m}\approx a_\ell^*a_m$.

Next, we consider the case $U=0.5$ where we expect similarities with the result depicted in the bottom panel in Fig.~\ref{fig1}. Indeed, it is seen in  Fig.~\ref{fig4}(b) that the number of bosons in the chain (which is given by ${\rm Tr}[\hat{\rho}]=\sum_n \lambda_n$) initially grows linearly with $\Delta$, however, for $\Delta\approx 1.0$ it drops back to zero. We also notice that for $U\ne0$ the system SPDM may differ from a pure state, i.e., $\lambda_2\ne0$.

Summarizing the obtained results, we come to the following intermediate conclusion.  One finds an excelent agreement between the classical and quantum approaches  in the case $U=0$ and a strong discrepancy in the case $U\ne0$. In the next section we quantify this discrepancy by using  the pseudoclassical approach.

\section{Pseudoclassical approach}
\label{sec5}

First, we clarify the meaning of the effective Planck constant entering Eq.~(\ref{a1}) and Eq.~(\ref{a2}).  It follows from these equations that the actual parameters, which determine the quantum dynamics, are $U'=\hbar U$ and $\Omega'=\Omega/\sqrt{\hbar}$. Remarkably, the  indicated scaling of the interaction constant and the Rabi frequency does not alter the classical dynamics of the system where we associate operators $\sqrt{\hbar}\hat{a}$ and  $\sqrt{\hbar}\hat{a}^\dagger$ with the canonical variables $a$ and $a^*$. Thus, the effective Planck constant $\hbar$ determines the mean number of bosons in the system through the relation $\bar{n}=|{\bf a}|^2/\hbar$. The lager is this number, the closer quantum system to its classical counterpart.  The pseudoclassical approach is an approximation to the exact quantum dynamics through series expansion in  the parameter $\hbar=1/\bar{n}$. It substitutes the master equation for the system density matrix by the Fokker-Planck  equation for the classical distribution function  $f=f({\bf a},{\bf a}^*,t)$ and, in this sense, is equivalent  to the truncated Wigner function approximation  \cite{Vogel89,Carusotto13,Kidd19} in the single-particle quantum mechanics.  Explicitly,  we have  \cite{bychek2020open}
\begin{eqnarray}
\nonumber
\frac{\partial f}{\partial t} = \left\{{\cal H},f\right\}\\  
\label{c1}
+ \frac{\gamma}{2}\left[\frac{\partial(a_L f)}{\partial a_L} + \frac{\partial(a_L^* f)}{\partial a_L^*}\right]
+ \frac{\hbar\gamma}{2}\frac{\partial^2 f}{\partial a_L \partial a_L^*} \;,
\end{eqnarray}
where
\begin{eqnarray}
\nonumber
{\cal H} =  \sum_{\ell=1}^L  \left[-(\Delta + \hbar U) |a_\ell |^2 + \frac{U}{2} |a_\ell |^4 \right]  \\
\label{c2}
- \frac{J}{2} \sum_{\ell}^L \Big(a^*_{\ell+1} a_\ell+ {\rm c.c.} \Big)
+\frac{\Omega}{2}(a_1 + a_1^*) \;,
\end{eqnarray}
and $\{\ldots,\ldots\}$ denote the Poisson brackets. 

Let us discuss the meaning of  different terms in the displayed equation.  The first term in the right-hand-side of this equation is the Liuoville equation for the conservative  chain. The second term describes the contraction  of the phase volume in the dissipative chain and, thus, can be referred to as friction. Finally, the last term describes  the diffusion. Using Eq.~(\ref{c1}) the SPDM is found as the phase-space average,
\begin{equation}
\label{cc}
\rho_{\ell,m}(t)=\int a_\ell^* a_m f({\bf a},{\bf a}^*,t) d{\bf a} d{\bf a}^*  \;.
\end{equation}
Usually, one evaluates the multi-dimensional integral in Eq.~(\ref{cc}) by putting into correspondence to the Fokker-Planck equation (\ref{c1}) the following Langevin equation,
\begin{equation}
\label{c3}
i\dot{a}_\ell=\frac{\partial H}{\partial a^*_\ell}
-i\frac{\gamma}{2} \delta_{\ell,L} a_\ell +\sqrt{\frac{\hbar \gamma}{2}}  \delta_{\ell,L} \xi(t) \;
\end{equation}
where $\xi(t)$ is the $\delta$-correlated white noise. Then the elements of SPDM are calculated as
\begin{equation}
\label{c4}
\rho_{\ell,m}(t)= \overline{a_\ell^*(t)a_m(t)} -\frac{1}{2} \delta_{\ell,m} \;,
\end{equation}
where the bar denotes the average over different realizations of the stochastic force $\xi(t)$.
\begin{figure}
\includegraphics[width=8.0cm,clip]{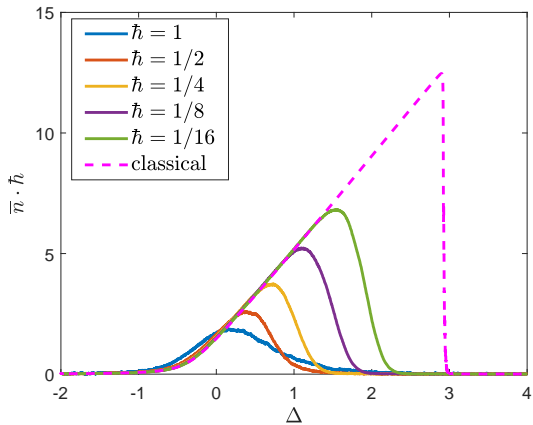}
\caption{The mean number of bosons in the dimer times the effective Planck constant according to the pseudoclassical approach (solid lines, average over 3600 realizations) for different values of the effective Planck constant. The other system parameters are  $J=0.5$, $U=0.5$, $\Omega=0.5$, $\gamma=0.2$, and the inverse sweeping rate $2\pi\cdot800/6$. The dashed line shows the classical result.}
\label{fig7}
\end{figure}

\subsection{Comparison with the exact results}

The primary advantage of the pseudoclassical approach as compared to the straightforward solution of the master equation is simplicity of numerical simulations which allows us to go deep in the semiclassical region. Of course, on the quantitative level, the pseudoclassical approach gives some systematic error. However, on the qualitative level, it correctly reproduces all main results of the quantum analysis. We illustrate this statement in the lower panel in Fig.~\ref{fig4} where we compare the SPDM  calculated by using the pseudoclassical approach (solid lines) with the exact result (dashed lines) for $\hbar=0.25$. It is seen in Fig.~\ref{fig4}(b) that the pseudoclassical approach correctly captures the decay of the SPDM long before $\Delta_{cr}$. In the next subsection we  use it to study this decay for the values of the effective Planck constant which are inaccessible in the exact quantum simulations. In fact, within  the pseudoclassical approach variation of the effective Planck constant affects only the noise intensity  while in the quantum equation of motion it rescales the inter-particle interaction and the amplitude of the driving force, which requires a proportional  increase of the truncation parameter $N_{max}$.

\subsection{Lifetime of the transporting state}

Quantum dynamics of the system calculated by using the pseudoclassical approach is exemplified in Fig.~\ref{fig7}. Shown are the mean number of bosons in the  chain $\bar{n}$, $\bar{n}={\rm Tr}[\hat{\rho}]=\sum_n \lambda_n$, times the effective Planck constant. It is seen that for $\hbar\rightarrow 0$ the quantum dynamics converges to the classical  result, where the destruction of the ballistic transport takes place  at $\Delta_{cr} \approx 3.0$.  The depicted in Fig.~\ref{fig7} results suggest the other critical detuning,
\begin{displaymath}
 \Delta_{qu}=\Delta_{qu}(\hbar,\beta) \le \Delta_{cr}  \;,
\end{displaymath}
at which the numbers of bosons in the chain is maximal. The fundamental reason for the inequality   $\Delta_{qu} \le \Delta_{cr}$ is the metastable character of the quantum attractor associated with the discussed transporting  state.
\begin{figure}[t]
\includegraphics[width=8.0cm,clip]{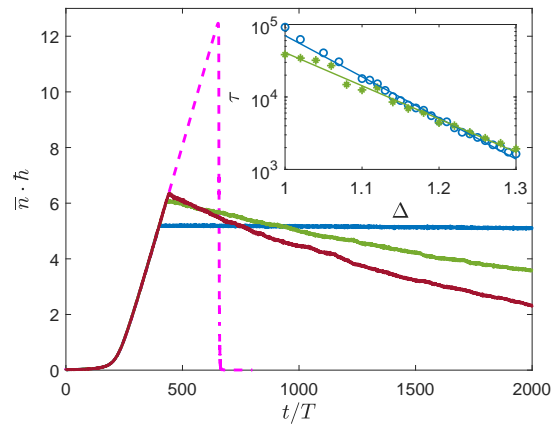}
\caption{Main panel: The mean number of bosons in the dimer times the effective Planck constant as a function of time for  $\Delta=1$, blue line, $\Delta=1.24$, green line, and $\Delta=1.30$, brown line. The value of the effective Planck constant $\hbar=1/16$.  The inset shows the lifetime of the transporting state as  the function of $\Delta$ for $L=2$ (open circles) and $L=8$ (asterisks).}
\label{fig8}
\end{figure}

To determine the lifetime $\tau=\tau(\Delta,\hbar)$ of the transporting state we evolve the system to $\Delta<\Delta_{qu}$ and then fix this detuning for the rest of time, see Fig.~\ref{fig8}. Then, by approximating the decay dynamics by the exponential function, we extract  $\tau$. The dependence of the lifetime $\tau$ on $\Delta$ is depicted in the inset in Fig.~\ref{fig8}. This result  suggests the following estimate for the lifetime,
\begin{equation}
\label{c5}
\tau\sim\exp\left(\frac{r}{\hbar}\right) \;,
\end{equation}
where $r=r(\Delta)$ is the basin size of the classical attractor.  Roughly, Eq.~(\ref{c5}) compares the minimal-size wave-packet with the basin size and, to insure the exponentially long lifetime of the considered transporting state, one should satisfy the condition $r(\Delta)\gg \hbar$.

\subsection{Long chain}
We repeated the above numerical simulations  for the chain of the length $L=8$ and obtained essentially the same results, see inset in Fig.~\ref{fig8}. The only new aspect is that for a long chain we can address the Anderson problem. It was found  that the discussed transporting state is insensitive to a weak on-site disorder $| \omega_\ell -\omega| \le \epsilon \ll \Delta$. One finds a qualitative expiation for this  result in terms of the  synchronization theory. In fact, the considered system of coupled nonlinear oscillators can be viewed as one  of physical realization of the Kuramoto model \cite{strogatz2000kuramoto}. The important property of the Kuramoto model is that  synchronization may occur for oscillators with different eigenfrequencies. In our case this means that the nonlinear oscillators  will be synchronized also in the presence of the on-site disorder, i.e., different linear frequencies $\omega_\ell$.

\section{Conclusion}

We study the transport of interacting Bose particles in the open Bose-Hubbard chain where the particles are injected in the first site of the chain and withdrawn from the last site. The analysis is done by using the pseudoclassical approximation which puts in correspondence to the open Bose-Hubbard model the chain of coupled nonlinear oscillators and where the transport of particles corresponds to the transport of excitations from the first to the last oscillator. It is shown that one can insure very efficient  transport of excitations by capturing the system into the classical attractor  which describes the synchronized oscillators. The quantum counterpart of this attractor corresponds to the quantum transporting state which, however, has a finite lifetime.  We obtain an estimate for the lifetime of this state and argue that it becomes exponentially long in the semiclassical limit.

\section{Acknowledgement}
This work has been supported by Russian Science Foundation through grant N19-12-00167.

\bibliography{mylib}

\providecommand{\noopsort}[1]{}\providecommand{\singleletter}[1]{#1}%
\begin{thebibliography}{25}%
\makeatletter
\providecommand \@ifxundefined [1]{%
 \@ifx{#1\undefined}
}%
\providecommand \@ifnum [1]{%
 \ifnum #1\expandafter \@firstoftwo
 \else \expandafter \@secondoftwo
 \fi
}%
\providecommand \@ifx [1]{%
 \ifx #1\expandafter \@firstoftwo
 \else \expandafter \@secondoftwo
 \fi
}%
\providecommand \natexlab [1]{#1}%
\providecommand \enquote  [1]{``#1''}%
\providecommand \bibnamefont  [1]{#1}%
\providecommand \bibfnamefont [1]{#1}%
\providecommand \citenamefont [1]{#1}%
\providecommand \href@noop [0]{\@secondoftwo}%
\providecommand \href [0]{\begingroup \@sanitize@url \@href}%
\providecommand \@href[1]{\@@startlink{#1}\@@href}%
\providecommand \@@href[1]{\endgroup#1\@@endlink}%
\providecommand \@sanitize@url [0]{\catcode `\\12\catcode `\$12\catcode
  `\&12\catcode `\#12\catcode `\^12\catcode `\_12\catcode `\%12\relax}%
\providecommand \@@startlink[1]{}%
\providecommand \@@endlink[0]{}%
\providecommand \url  [0]{\begingroup\@sanitize@url \@url }%
\providecommand \@url [1]{\endgroup\@href {#1}{\urlprefix }}%
\providecommand \urlprefix  [0]{URL }%
\providecommand \Eprint [0]{\href }%
\providecommand \doibase [0]{http://dx.doi.org/}%
\providecommand \selectlanguage [0]{\@gobble}%
\providecommand \bibinfo  [0]{\@secondoftwo}%
\providecommand \bibfield  [0]{\@secondoftwo}%
\providecommand \translation [1]{[#1]}%
\providecommand \BibitemOpen [0]{}%
\providecommand \bibitemStop [0]{}%
\providecommand \bibitemNoStop [0]{.\EOS\space}%
\providecommand \EOS [0]{\spacefactor3000\relax}%
\providecommand \BibitemShut  [1]{\csname bibitem#1\endcsname}%
\let\auto@bib@innerbib\@empty
\bibitem [{\citenamefont {Ivanov}\ \emph {et~al.}(2013)\citenamefont {Ivanov},
  \citenamefont {Kordas}, \citenamefont {Komnik},\ and\ \citenamefont
  {Wimberger}}]{iv2013bosonic}%
  \BibitemOpen
  \bibfield  {author} {\bibinfo {author} {\bibfnamefont {A.}~\bibnamefont
  {Ivanov}}, \bibinfo {author} {\bibfnamefont {G.}~\bibnamefont {Kordas}},
  \bibinfo {author} {\bibfnamefont {A.}~\bibnamefont {Komnik}}, \ and\ \bibinfo
  {author} {\bibfnamefont {S.}~\bibnamefont {Wimberger}},\ }\bibfield  {title}
  {\enquote {\bibinfo {title} {Bosonic transport through a chain of quantum
  dots},}\ }\href@noop {} {\bibfield  {journal} {\bibinfo  {journal} {The
  European Physical Journal B}\ }\textbf {\bibinfo {volume} {86}},\ \bibinfo
  {pages} {1--7} (\bibinfo {year} {2013})}\BibitemShut {NoStop}%
\bibitem [{\citenamefont {Kordas}\ \emph {et~al.}(2015)\citenamefont {Kordas},
  \citenamefont {Witthaut}, \citenamefont {Buonsante}, \citenamefont {Vezzani},
  \citenamefont {Burioni}, \citenamefont {Karanikas},\ and\ \citenamefont
  {Wimberger}}]{kordas2015dissipative}%
  \BibitemOpen
  \bibfield  {author} {\bibinfo {author} {\bibfnamefont {G.}~\bibnamefont
  {Kordas}}, \bibinfo {author} {\bibfnamefont {D.}~\bibnamefont {Witthaut}},
  \bibinfo {author} {\bibfnamefont {P.}~\bibnamefont {Buonsante}}, \bibinfo
  {author} {\bibfnamefont {A.}~\bibnamefont {Vezzani}}, \bibinfo {author}
  {\bibfnamefont {R.}~\bibnamefont {Burioni}}, \bibinfo {author} {\bibfnamefont
  {A.I.}\ \bibnamefont {Karanikas}}, \ and\ \bibinfo {author} {\bibfnamefont
  {S.}~\bibnamefont {Wimberger}},\ }\bibfield  {title} {\enquote {\bibinfo
  {title} {The dissipative {Bose-Hubbard} model},}\ }\href@noop {} {\bibfield
  {journal} {\bibinfo  {journal} {The European Physical Journal Special
  Topics}\ }\textbf {\bibinfo {volume} {224}},\ \bibinfo {pages} {2127--2171}
  (\bibinfo {year} {2015})}\BibitemShut {NoStop}%
\bibitem [{\citenamefont {Kolovsky}\ \emph {et~al.}(2018)\citenamefont
  {Kolovsky}, \citenamefont {Denis},\ and\ \citenamefont
  {Wimberger}}]{kolovsky2018landauer}%
  \BibitemOpen
  \bibfield  {author} {\bibinfo {author} {\bibfnamefont {A.R.}\ \bibnamefont
  {Kolovsky}}, \bibinfo {author} {\bibfnamefont {Z.}~\bibnamefont {Denis}}, \
  and\ \bibinfo {author} {\bibfnamefont {S.}~\bibnamefont {Wimberger}},\
  }\bibfield  {title} {\enquote {\bibinfo {title} {{Landauer-B{\"u}ttiker}
  equation for bosonic carriers},}\ }\href@noop {} {\bibfield  {journal}
  {\bibinfo  {journal} {Physical Review A}\ }\textbf {\bibinfo {volume} {98}},\
  \bibinfo {pages} {043623} (\bibinfo {year} {2018})}\BibitemShut {NoStop}%
\bibitem [{\citenamefont {Bychek}\ \emph {et~al.}(2020)\citenamefont {Bychek},
  \citenamefont {Muraev}, \citenamefont {Maksimov},\ and\ \citenamefont
  {Kolovsky}}]{bychek2020open}%
  \BibitemOpen
  \bibfield  {author} {\bibinfo {author} {\bibfnamefont {A.~A.}\ \bibnamefont
  {Bychek}}, \bibinfo {author} {\bibfnamefont {P.~S.}\ \bibnamefont {Muraev}},
  \bibinfo {author} {\bibfnamefont {D.~N.}\ \bibnamefont {Maksimov}}, \ and\
  \bibinfo {author} {\bibfnamefont {A.~R.}\ \bibnamefont {Kolovsky}},\
  }\bibfield  {title} {\enquote {\bibinfo {title} {Open {Bose-Hubbard} chain:
  Pseudoclassical approach},}\ }\href@noop {} {\bibfield  {journal} {\bibinfo
  {journal} {Physical Review E}\ }\textbf {\bibinfo {volume} {101}},\ \bibinfo
  {pages} {012208} (\bibinfo {year} {2020})}\BibitemShut {NoStop}%
\bibitem [{\citenamefont {Muraev}\ \emph {et~al.}(2022)\citenamefont {Muraev},
  \citenamefont {Maksimov},\ and\ \citenamefont
  {Kolovsky}}]{muraev2022resonant}%
  \BibitemOpen
  \bibfield  {author} {\bibinfo {author} {\bibfnamefont {P.~S.}\ \bibnamefont
  {Muraev}}, \bibinfo {author} {\bibfnamefont {D.~N.}\ \bibnamefont
  {Maksimov}}, \ and\ \bibinfo {author} {\bibfnamefont {A.~R.}\ \bibnamefont
  {Kolovsky}},\ }\bibfield  {title} {\enquote {\bibinfo {title} {Resonant
  transport of bosonic carriers through a quantum device},}\ }\href@noop {}
  {\bibfield  {journal} {\bibinfo  {journal} {Physical Review A}\ }\textbf
  {\bibinfo {volume} {105}},\ \bibinfo {pages} {013307} (\bibinfo {year}
  {2022})}\BibitemShut {NoStop}%
\bibitem [{\citenamefont {Raftery}\ \emph {et~al.}(2014)\citenamefont
  {Raftery}, \citenamefont {Sadri}, \citenamefont {Schmidt}, \citenamefont
  {T{\"u}reci},\ and\ \citenamefont {Houck}}]{raftery2014observation}%
  \BibitemOpen
  \bibfield  {author} {\bibinfo {author} {\bibfnamefont {J.}~\bibnamefont
  {Raftery}}, \bibinfo {author} {\bibfnamefont {D.}~\bibnamefont {Sadri}},
  \bibinfo {author} {\bibfnamefont {S.}~\bibnamefont {Schmidt}}, \bibinfo
  {author} {\bibfnamefont {H.E.}\ \bibnamefont {T{\"u}reci}}, \ and\ \bibinfo
  {author} {\bibfnamefont {A.A.}\ \bibnamefont {Houck}},\ }\bibfield  {title}
  {\enquote {\bibinfo {title} {Observation of a dissipation-induced classical
  to quantum transition},}\ }\href@noop {} {\bibfield  {journal} {\bibinfo
  {journal} {Physical Review X}\ }\textbf {\bibinfo {volume} {4}},\ \bibinfo
  {pages} {031043} (\bibinfo {year} {2014})}\BibitemShut {NoStop}%
\bibitem [{\citenamefont {Fitzpatrick}\ \emph {et~al.}(2017)\citenamefont
  {Fitzpatrick}, \citenamefont {Sundaresan}, \citenamefont {Li}, \citenamefont
  {Koch},\ and\ \citenamefont {Houck}}]{fitzpatrick2017observation}%
  \BibitemOpen
  \bibfield  {author} {\bibinfo {author} {\bibfnamefont {M.}~\bibnamefont
  {Fitzpatrick}}, \bibinfo {author} {\bibfnamefont {N.M.}\ \bibnamefont
  {Sundaresan}}, \bibinfo {author} {\bibfnamefont {A.C.Y.}\ \bibnamefont {Li}},
  \bibinfo {author} {\bibfnamefont {J.}~\bibnamefont {Koch}}, \ and\ \bibinfo
  {author} {\bibfnamefont {A.A.}\ \bibnamefont {Houck}},\ }\bibfield  {title}
  {\enquote {\bibinfo {title} {Observation of a dissipative phase transition in
  a one-dimensional circuit {QED} lattice},}\ }\href@noop {} {\bibfield
  {journal} {\bibinfo  {journal} {Physical Review X}\ }\textbf {\bibinfo
  {volume} {7}},\ \bibinfo {pages} {011016} (\bibinfo {year}
  {2017})}\BibitemShut {NoStop}%
\bibitem [{\citenamefont {Fedorov}\ \emph {et~al.}(2021)\citenamefont
  {Fedorov}, \citenamefont {Remizov}, \citenamefont {Shapiro}, \citenamefont
  {Pogosov}, \citenamefont {Egorova}, \citenamefont {Tsitsilin}, \citenamefont
  {Andronik}, \citenamefont {Dobronosova}, \citenamefont {Rodionov},
  \citenamefont {Astafiev},\ and\ \citenamefont {Ustinov}}]{fedorov2021photon}%
  \BibitemOpen
  \bibfield  {author} {\bibinfo {author} {\bibfnamefont {G.P.}\ \bibnamefont
  {Fedorov}}, \bibinfo {author} {\bibfnamefont {S.V.}\ \bibnamefont {Remizov}},
  \bibinfo {author} {\bibfnamefont {D.S.}\ \bibnamefont {Shapiro}}, \bibinfo
  {author} {\bibfnamefont {W.V.}\ \bibnamefont {Pogosov}}, \bibinfo {author}
  {\bibfnamefont {E.}~\bibnamefont {Egorova}}, \bibinfo {author} {\bibfnamefont
  {I.}~\bibnamefont {Tsitsilin}}, \bibinfo {author} {\bibfnamefont
  {M.}~\bibnamefont {Andronik}}, \bibinfo {author} {\bibfnamefont {A.A.}\
  \bibnamefont {Dobronosova}}, \bibinfo {author} {\bibfnamefont {I.A.}\
  \bibnamefont {Rodionov}}, \bibinfo {author} {\bibfnamefont {O.V.}\
  \bibnamefont {Astafiev}}, \ and\ \bibinfo {author} {\bibfnamefont {A.V.}\
  \bibnamefont {Ustinov}},\ }\bibfield  {title} {\enquote {\bibinfo {title}
  {Photon transport in a {Bose-Hubbard} chain of superconducting artificial
  atoms},}\ }\href@noop {} {\bibfield  {journal} {\bibinfo  {journal} {Physical
  Review Letters}\ }\textbf {\bibinfo {volume} {126}},\ \bibinfo {pages}
  {180503} (\bibinfo {year} {2021})}\BibitemShut {NoStop}%
\bibitem [{\citenamefont {Lagoudakis}\ \emph {et~al.}(2010)\citenamefont
  {Lagoudakis}, \citenamefont {Pietka}, \citenamefont {Wouters}, \citenamefont
  {Andr{\'e}},\ and\ \citenamefont
  {Deveaud-Pl{\'e}dran}}]{lagoudakis2010coherent}%
  \BibitemOpen
  \bibfield  {author} {\bibinfo {author} {\bibfnamefont {K.G.}\ \bibnamefont
  {Lagoudakis}}, \bibinfo {author} {\bibfnamefont {B.}~\bibnamefont {Pietka}},
  \bibinfo {author} {\bibfnamefont {M.}~\bibnamefont {Wouters}}, \bibinfo
  {author} {\bibfnamefont {R.}~\bibnamefont {Andr{\'e}}}, \ and\ \bibinfo
  {author} {\bibfnamefont {B.}~\bibnamefont {Deveaud-Pl{\'e}dran}},\ }\bibfield
   {title} {\enquote {\bibinfo {title} {Coherent oscillations in an
  exciton-polariton {Josephson} junction},}\ }\href@noop {} {\bibfield
  {journal} {\bibinfo  {journal} {Physical review letters}\ }\textbf {\bibinfo
  {volume} {105}},\ \bibinfo {pages} {120403} (\bibinfo {year}
  {2010})}\BibitemShut {NoStop}%
\bibitem [{\citenamefont {Abbarchi}\ \emph {et~al.}(2013)\citenamefont
  {Abbarchi}, \citenamefont {Amo}, \citenamefont {Sala}, \citenamefont
  {Solnyshkov}, \citenamefont {Flayac}, \citenamefont {Ferrier}, \citenamefont
  {Sagnes}, \citenamefont {Galopin}, \citenamefont {Lema{\^\i}tre},
  \citenamefont {Malpuech},\ and\ \citenamefont
  {Bloch}}]{abbarchi2013macroscopic}%
  \BibitemOpen
  \bibfield  {author} {\bibinfo {author} {\bibfnamefont {M.}~\bibnamefont
  {Abbarchi}}, \bibinfo {author} {\bibfnamefont {A.}~\bibnamefont {Amo}},
  \bibinfo {author} {\bibfnamefont {V.G.}\ \bibnamefont {Sala}}, \bibinfo
  {author} {\bibfnamefont {D.D.}\ \bibnamefont {Solnyshkov}}, \bibinfo {author}
  {\bibfnamefont {H.}~\bibnamefont {Flayac}}, \bibinfo {author} {\bibfnamefont
  {L.}~\bibnamefont {Ferrier}}, \bibinfo {author} {\bibfnamefont
  {I.}~\bibnamefont {Sagnes}}, \bibinfo {author} {\bibfnamefont
  {E.}~\bibnamefont {Galopin}}, \bibinfo {author} {\bibfnamefont
  {A.}~\bibnamefont {Lema{\^\i}tre}}, \bibinfo {author} {\bibfnamefont
  {G.}~\bibnamefont {Malpuech}}, \ and\ \bibinfo {author} {\bibfnamefont
  {J.}~\bibnamefont {Bloch}},\ }\bibfield  {title} {\enquote {\bibinfo {title}
  {Macroscopic quantum self-trapping and {Josephson} oscillations of exciton
  polaritons},}\ }\href@noop {} {\bibfield  {journal} {\bibinfo  {journal}
  {Nature Physics}\ }\textbf {\bibinfo {volume} {9}},\ \bibinfo {pages}
  {275--279} (\bibinfo {year} {2013})}\BibitemShut {NoStop}%
\bibitem [{\citenamefont {Cao}\ \emph {et~al.}(2016)\citenamefont {Cao},
  \citenamefont {Mahmud},\ and\ \citenamefont {Hafezi}}]{cao2016two}%
  \BibitemOpen
  \bibfield  {author} {\bibinfo {author} {\bibfnamefont {Bin}\ \bibnamefont
  {Cao}}, \bibinfo {author} {\bibfnamefont {K.W.}\ \bibnamefont {Mahmud}}, \
  and\ \bibinfo {author} {\bibfnamefont {M.}~\bibnamefont {Hafezi}},\
  }\bibfield  {title} {\enquote {\bibinfo {title} {Two coupled nonlinear
  cavities in a driven-dissipative environment},}\ }\href@noop {} {\bibfield
  {journal} {\bibinfo  {journal} {Physical Review A}\ }\textbf {\bibinfo
  {volume} {94}},\ \bibinfo {pages} {063805} (\bibinfo {year}
  {2016})}\BibitemShut {NoStop}%
\bibitem [{\citenamefont {Casteels}\ and\ \citenamefont
  {Ciuti}(2017)}]{casteels2017quantum}%
  \BibitemOpen
  \bibfield  {author} {\bibinfo {author} {\bibfnamefont {W.}~\bibnamefont
  {Casteels}}\ and\ \bibinfo {author} {\bibfnamefont {C.}~\bibnamefont
  {Ciuti}},\ }\bibfield  {title} {\enquote {\bibinfo {title} {Quantum
  entanglement in the spatial-symmetry-breaking phase transition of a
  driven-dissipative {Bose-Hubbard} dimer},}\ }\href@noop {} {\bibfield
  {journal} {\bibinfo  {journal} {Physical Review A}\ }\textbf {\bibinfo
  {volume} {95}},\ \bibinfo {pages} {013812} (\bibinfo {year}
  {2017})}\BibitemShut {NoStop}%
\bibitem [{\citenamefont {Rodriguez}\ \emph {et~al.}(2017)\citenamefont
  {Rodriguez}, \citenamefont {Casteels}, \citenamefont {Storme}, \citenamefont
  {Zambon}, \citenamefont {Sagnes}, \citenamefont {{Le Gratiet}}, \citenamefont
  {Galopin}, \citenamefont {Lema{\^\i}tre}, \citenamefont {Amo}, \citenamefont
  {Ciuti},\ and\ \citenamefont {Bloch}}]{rodriguez2017probing}%
  \BibitemOpen
  \bibfield  {author} {\bibinfo {author} {\bibfnamefont {S.R.K.}\ \bibnamefont
  {Rodriguez}}, \bibinfo {author} {\bibfnamefont {W.}~\bibnamefont {Casteels}},
  \bibinfo {author} {\bibfnamefont {F.}~\bibnamefont {Storme}}, \bibinfo
  {author} {\bibfnamefont {N.C.}\ \bibnamefont {Zambon}}, \bibinfo {author}
  {\bibfnamefont {I.}~\bibnamefont {Sagnes}}, \bibinfo {author} {\bibfnamefont
  {L.}~\bibnamefont {{Le Gratiet}}}, \bibinfo {author} {\bibfnamefont
  {E.}~\bibnamefont {Galopin}}, \bibinfo {author} {\bibfnamefont
  {A.}~\bibnamefont {Lema{\^\i}tre}}, \bibinfo {author} {\bibfnamefont
  {A.}~\bibnamefont {Amo}}, \bibinfo {author} {\bibfnamefont {C.}~\bibnamefont
  {Ciuti}}, \ and\ \bibinfo {author} {\bibfnamefont {J.}~\bibnamefont
  {Bloch}},\ }\bibfield  {title} {\enquote {\bibinfo {title} {Probing a
  dissipative phase transition via dynamical optical hysteresis},}\ }\href@noop
  {} {\bibfield  {journal} {\bibinfo  {journal} {Physical review letters}\
  }\textbf {\bibinfo {volume} {118}},\ \bibinfo {pages} {247402} (\bibinfo
  {year} {2017})}\BibitemShut {NoStop}%
\bibitem [{\citenamefont {Lled{\'o}}\ and\ \citenamefont
  {Szyma{\'n}ska}(2020)}]{lledo2020dissipative}%
  \BibitemOpen
  \bibfield  {author} {\bibinfo {author} {\bibfnamefont {C.}~\bibnamefont
  {Lled{\'o}}}\ and\ \bibinfo {author} {\bibfnamefont {M.H.}\ \bibnamefont
  {Szyma{\'n}ska}},\ }\bibfield  {title} {\enquote {\bibinfo {title} {A
  dissipative time crystal with or without $\mathrm{Z}_2$ symmetry breaking},}\
  }\href@noop {} {\bibfield  {journal} {\bibinfo  {journal} {New Journal of
  Physics}\ }\textbf {\bibinfo {volume} {22}},\ \bibinfo {pages} {075002}
  (\bibinfo {year} {2020})}\BibitemShut {NoStop}%
\bibitem [{\citenamefont {Zambon}\ \emph {et~al.}(2020)\citenamefont {Zambon},
  \citenamefont {Rodriguez}, \citenamefont {Lema{\^\i}tre}, \citenamefont
  {Harouri}, \citenamefont {{Le Gratiet}}, \citenamefont {Sagnes},
  \citenamefont {{St-Jean}}, \citenamefont {Ravets}, \citenamefont {Amo},\ and\
  \citenamefont {Bloch}}]{zambon2020parametric}%
  \BibitemOpen
  \bibfield  {author} {\bibinfo {author} {\bibfnamefont {N.C.}\ \bibnamefont
  {Zambon}}, \bibinfo {author} {\bibfnamefont {S.R.K.}\ \bibnamefont
  {Rodriguez}}, \bibinfo {author} {\bibfnamefont {A.}~\bibnamefont
  {Lema{\^\i}tre}}, \bibinfo {author} {\bibfnamefont {A.}~\bibnamefont
  {Harouri}}, \bibinfo {author} {\bibfnamefont {L.}~\bibnamefont {{Le
  Gratiet}}}, \bibinfo {author} {\bibfnamefont {I.}~\bibnamefont {Sagnes}},
  \bibinfo {author} {\bibfnamefont {P.}~\bibnamefont {{St-Jean}}}, \bibinfo
  {author} {\bibfnamefont {S.}~\bibnamefont {Ravets}}, \bibinfo {author}
  {\bibfnamefont {A.}~\bibnamefont {Amo}}, \ and\ \bibinfo {author}
  {\bibfnamefont {J.}~\bibnamefont {Bloch}},\ }\bibfield  {title} {\enquote
  {\bibinfo {title} {Parametric instability in coupled nonlinear
  microcavities},}\ }\href@noop {} {\bibfield  {journal} {\bibinfo  {journal}
  {Physical Review A}\ }\textbf {\bibinfo {volume} {102}},\ \bibinfo {pages}
  {023526} (\bibinfo {year} {2020})}\BibitemShut {NoStop}%
\bibitem [{\citenamefont {Astrakharchik}\ and\ \citenamefont
  {Pitaevskii}(2004)}]{astrakharchik2004motion}%
  \BibitemOpen
  \bibfield  {author} {\bibinfo {author} {\bibfnamefont {G.E.}\ \bibnamefont
  {Astrakharchik}}\ and\ \bibinfo {author} {\bibfnamefont {L.P.}\ \bibnamefont
  {Pitaevskii}},\ }\bibfield  {title} {\enquote {\bibinfo {title} {Motion of a
  heavy impurity through a {Bose-Einstein} condensate},}\ }\href@noop {}
  {\bibfield  {journal} {\bibinfo  {journal} {Physical Review A}\ }\textbf
  {\bibinfo {volume} {70}},\ \bibinfo {pages} {013608} (\bibinfo {year}
  {2004})}\BibitemShut {NoStop}%
\bibitem [{\citenamefont {Cherny}\ \emph {et~al.}(2012)\citenamefont {Cherny},
  \citenamefont {Caux},\ and\ \citenamefont {Brand}}]{cherny2012theory}%
  \BibitemOpen
  \bibfield  {author} {\bibinfo {author} {\bibfnamefont {A.~Yu.}\ \bibnamefont
  {Cherny}}, \bibinfo {author} {\bibfnamefont {J.-S.}\ \bibnamefont {Caux}}, \
  and\ \bibinfo {author} {\bibfnamefont {J.}~\bibnamefont {Brand}},\ }\bibfield
   {title} {\enquote {\bibinfo {title} {Theory of superfluidity and drag force
  in the one-dimensional {Bose gas}},}\ }\href@noop {} {\bibfield  {journal}
  {\bibinfo  {journal} {Frontiers of Physics}\ }\textbf {\bibinfo {volume}
  {7}},\ \bibinfo {pages} {54--71} (\bibinfo {year} {2012})}\BibitemShut
  {NoStop}%
\bibitem [{\citenamefont {Giraldo}\ \emph {et~al.}(2022)\citenamefont
  {Giraldo}, \citenamefont {Masson}, \citenamefont {Broderick},\ and\
  \citenamefont {Krauskopf}}]{giraldo2022semiclassical}%
  \BibitemOpen
  \bibfield  {author} {\bibinfo {author} {\bibfnamefont {A.}~\bibnamefont
  {Giraldo}}, \bibinfo {author} {\bibfnamefont {S.J.}\ \bibnamefont {Masson}},
  \bibinfo {author} {\bibfnamefont {N.G.R.}\ \bibnamefont {Broderick}}, \ and\
  \bibinfo {author} {\bibfnamefont {B.}~\bibnamefont {Krauskopf}},\ }\bibfield
  {title} {\enquote {\bibinfo {title} {Semiclassical bifurcations and quantum
  trajectories: a case study of the open {Bose--Hubbard} dimer},}\ }\href@noop
  {} {\bibfield  {journal} {\bibinfo  {journal} {The European Physical Journal
  Special Topics}\ ,\ \bibinfo {pages} {1--17}} (\bibinfo {year}
  {2022})}\BibitemShut {NoStop}%
\bibitem [{\citenamefont {Lichtenberg}\ and\ \citenamefont
  {Lieberman}(2013)}]{lichtenberg2013regular}%
  \BibitemOpen
  \bibfield  {author} {\bibinfo {author} {\bibfnamefont {A.J.}\ \bibnamefont
  {Lichtenberg}}\ and\ \bibinfo {author} {\bibfnamefont {M.A.}\ \bibnamefont
  {Lieberman}},\ }\href@noop {} {\emph {\bibinfo {title} {Regular and
  stochastic motion}}},\ Vol.~\bibinfo {volume} {38}\ (\bibinfo  {publisher}
  {Springer Science \& Business Media},\ \bibinfo {year} {2013})\BibitemShut
  {NoStop}%
\bibitem [{\citenamefont {Landau}\ and\ \citenamefont
  {Lifshitz}(1976)}]{landau1976mechanics}%
  \BibitemOpen
  \bibfield  {author} {\bibinfo {author} {\bibfnamefont {L.D.}\ \bibnamefont
  {Landau}}\ and\ \bibinfo {author} {\bibfnamefont {E.M.}\ \bibnamefont
  {Lifshitz}},\ }\bibfield  {title} {\enquote {\bibinfo {title} {Mechanics},}\
  }\href@noop {} {\bibfield  {journal} {\bibinfo  {journal} {New York}\ ,\
  \bibinfo {pages} {93}} (\bibinfo {year} {1976})}\BibitemShut {NoStop}%
\bibitem [{\citenamefont {Shrimali}\ \emph {et~al.}(2008)\citenamefont
  {Shrimali}, \citenamefont {Prasad}, \citenamefont {Ramaswamy},\ and\
  \citenamefont {Feudel}}]{shrimali2008nature}%
  \BibitemOpen
  \bibfield  {author} {\bibinfo {author} {\bibfnamefont {M.D.}\ \bibnamefont
  {Shrimali}}, \bibinfo {author} {\bibfnamefont {A.}~\bibnamefont {Prasad}},
  \bibinfo {author} {\bibfnamefont {R.}~\bibnamefont {Ramaswamy}}, \ and\
  \bibinfo {author} {\bibfnamefont {U.}~\bibnamefont {Feudel}},\ }\bibfield
  {title} {\enquote {\bibinfo {title} {The nature of attractor basins in
  multistable systems},}\ }\href@noop {} {\bibfield  {journal} {\bibinfo
  {journal} {International Journal of Bifurcation and Chaos}\ }\textbf
  {\bibinfo {volume} {18}},\ \bibinfo {pages} {1675--1688} (\bibinfo {year}
  {2008})}\BibitemShut {NoStop}%
\bibitem [{\citenamefont {Vogel}\ and\ \citenamefont {Risken}(1989)}]{Vogel89}%
  \BibitemOpen
  \bibfield  {author} {\bibinfo {author} {\bibfnamefont {K.}~\bibnamefont
  {Vogel}}\ and\ \bibinfo {author} {\bibfnamefont {H.}~\bibnamefont {Risken}},\
  }\bibfield  {title} {\enquote {\bibinfo {title} {Quasiprobability
  distributions in dispersive optical bistability},}\ }\href {\doibase
  10.1103/physreva.39.4675} {\bibfield  {journal} {\bibinfo  {journal}
  {Physical Review A}\ }\textbf {\bibinfo {volume} {39}},\ \bibinfo {pages}
  {4675--4683} (\bibinfo {year} {1989})}\BibitemShut {NoStop}%
\bibitem [{\citenamefont {C.}\ and\ \citenamefont {C.}(2013)}]{Carusotto13}%
  \BibitemOpen
  \bibfield  {author} {\bibinfo {author} {\bibfnamefont {Iacopo}\ \bibnamefont
  {C.}}\ and\ \bibinfo {author} {\bibfnamefont {Cristiano}\ \bibnamefont
  {C.}},\ }\bibfield  {title} {\enquote {\bibinfo {title} {Quantum fluids of
  light},}\ }\href {\doibase 10.1103/revmodphys.85.299} {\bibfield  {journal}
  {\bibinfo  {journal} {Reviews of Modern Physics}\ }\textbf {\bibinfo {volume}
  {85}},\ \bibinfo {pages} {299--366} (\bibinfo {year} {2013})}\BibitemShut
  {NoStop}%
\bibitem [{\citenamefont {Kidd}\ \emph {et~al.}(2019)\citenamefont {Kidd},
  \citenamefont {Olsen},\ and\ \citenamefont {Corney}}]{Kidd19}%
  \BibitemOpen
  \bibfield  {author} {\bibinfo {author} {\bibfnamefont {R.A.}\ \bibnamefont
  {Kidd}}, \bibinfo {author} {\bibfnamefont {M.K.}\ \bibnamefont {Olsen}}, \
  and\ \bibinfo {author} {\bibfnamefont {J.F.}\ \bibnamefont {Corney}},\
  }\bibfield  {title} {\enquote {\bibinfo {title} {Quantum chaos in a
  {Bose-Hubbard} dimer with modulated tunneling},}\ }\href {\doibase
  10.1103/physreva.100.013625} {\bibfield  {journal} {\bibinfo  {journal}
  {Physical Review A}\ }\textbf {\bibinfo {volume} {100}} (\bibinfo {year}
  {2019}),\ 10.1103/physreva.100.013625}\BibitemShut {NoStop}%
\bibitem [{\citenamefont {Strogatz}(2000)}]{strogatz2000kuramoto}%
  \BibitemOpen
  \bibfield  {author} {\bibinfo {author} {\bibfnamefont {S.H.}\ \bibnamefont
  {Strogatz}},\ }\bibfield  {title} {\enquote {\bibinfo {title} {From {Kuramoto
  to Crawford}: exploring the onset of synchronization in populations of
  coupled oscillators},}\ }\href@noop {} {\bibfield  {journal} {\bibinfo
  {journal} {Physica D: Nonlinear Phenomena}\ }\textbf {\bibinfo {volume}
  {143}},\ \bibinfo {pages} {1--20} (\bibinfo {year} {2000})}\BibitemShut
  {NoStop}%
\end{thebibliography}%
\end{document}